\documentclass[conference]{IEEEtran}

\IEEEoverridecommandlockouts
\usepackage{cite}
\usepackage{amsmath,amssymb,amsfonts}
\usepackage{algorithmic}
\usepackage{graphicx}
\usepackage{textcomp}
\usepackage{xcolor}
\usepackage{algorithm}
\usepackage{algorithmic}
\usepackage{url}
\usepackage{bm}

\def\BibTeX{{\rm B\kern-.05em{\sc i\kern-.025em b}\kern-.08em
    T\kern-.1667em\lower.7ex\hbox{E}\kern-.125emX}}
\begin{document}

\title{Soft-Output Deep Neural Network-Based Decoding}

\author{
  \IEEEauthorblockN{Dmitry Artemasov, Kirill Andreev, Pavel Rybin, Alexey Frolov}

  \IEEEauthorblockA{\textit{Center for Next Generation Wireless and IoT} \\
  \textit{Skolkovo Institute of Science and Technology}\\
  Moscow, Russia \\
  \{dmitry.artemasov, k.andreev, p.rybin, al.frolov\}@skoltech.ru}

  \thanks{The research was carried at Skolkovo Institute of Science and Technology and supported by the Russian Science Foundation (project no. 23-11-00340), \protect\url{https://rscf.ru/en/project/23-11-00340/}}
}

\maketitle

\begin{abstract}
Deep neural network (DNN)-based channel decoding is widely considered in the literature. The existing solutions are investigated for the case of hard output, i.e. when the decoder returns the estimated information word. At the same time, soft-output decoding is of critical importance for iterative receivers and decoders. In this paper, we focus on the soft-output DNN-based decoding problem. We start with the syndrome-based approach proposed by Bennatan et al. (2018) and modify it to provide soft output in the AWGN channel. The new decoder can be considered as an approximation of the MAP decoder with smaller computation complexity. We discuss various regularization functions for joint DNN-MAP training and compare the resulting distributions for $[64, 45]$ BCH code. Finally, to demonstrate the soft-output quality we consider the turbo-product code with $[64, 45]$ BCH codes as row and column codes. We show that the resulting DNN-based scheme is very close to the MAP-based performance and significantly outperforms the solution based on the Chase decoder. We come to the conclusion that the new method is prospective for the challenging problem of DNN-based decoding of long codes consisting of short component codes.
\end{abstract}


\section{Introduction}

Nowadays, the scope of application of machine learning algorithms and deep neural networks (DNN) is growing rapidly. In the past decade, the use of DNNs has allowed groundbreaking results to be achieved in applications such as image, video, and natural language processing \cite{DeepLearningBook}. All of these applications deal with natural signals. At the same time, much less attention has been devoted to the application of ML methods in communications. In this paper, we consider the application of ML algorithms for the channel decoding problem. To justify this research direction we note that the decoding problem is a classification problem: the channel output must correspond to one of the classes (codewords). The significant difference between this problem and a typical classification problem lies in the exponentially large number of classes.


The idea to use NNs in the channel decoding problem is not new, here we mention the early papers \cite{neural-viterbi, Tallini95}. But due to the lack of computation capabilities, these methods were forgotten until a recent paper \cite{TenBrink2017a}. The authors of \cite{TenBrink2017a} consider a binary input channel with additive white Gaussian noise (AWGN) and utilize a fully connected NN as a decoder. The major ML challenge is dataset collection and labeling but in the decoding task, this problem disappears as a dataset of any size can be generated easily. At the same time, the approach of~\cite{TenBrink2017a} suffers from the ``curse of dimensionality'' problem, as the number of codewords is exponential in the number of information bits. Thus, for any reasonable parameters, it is not possible to train the NN on all the codewords. The only hope is that the NN can learn the code structure by observing a small number of codewords. Note that all the practical codes are linear ones and can be defined by a basis, so the basis vectors are sufficient to learn the code structure. The main outcome of \cite{TenBrink2017a} is that the fully connected NN cannot learn the code structure\footnote{we note that for structured codes, such as linear codes, the NN can work on some codewords that were not shown to it, see \cite{TenBrink2017a} for more details.} and thus the such method is applicable for very short codes only. The subsequent articles propose to combine the existing decoding algorithms and NNs. The articles \cite{Burshtein2018, LugoschG2017, 5g-short-bp, PoorPLDPC2021, Pfister2019, nachmani2017rnn, Vasic2018} consider belief propagation algorithm (both Sum-Product and Min-Sum modifications) which is suitable for any linear code, but shows the best results for sparse-graph codes, such as Low-Density Parity-Check (LDPC) codes \cite{Gallager63}. The idea is to unwrap (or unroll) the underlying Tanner graph and obtain a sparse NN, which repeats the decoder operations but is equipped with trainable weights. The improvements were obtained for BCH codes \cite{Burshtein2018, LugoschG2017, nachmani2017rnn} and LDPC codes \cite{5g-short-bp, PoorPLDPC2021, Vasic2018}. The next idea was to replace the activation functions, the architecture is called a hyper-network \cite{Nachmani2019, Nachmani2020}. Later, Cammerer et al. proposed to replace node and edge message updates with trainable functions, thus allowing NN to learn a generalized message passing algorithm \cite{bp-cammerer-2022}. Another approach proposed in \cite{Bennatan2018} is to consider the syndrome-based decoding algorithm that is suitable for any linear codes. The basic syndrome-based decoding algorithm implies the use of the mapping (syndrome to the coset leader), which has the exponential (in the number of parity-check bits) size. The idea of \cite{Bennatan2018} is to approximate this table with a NN. We note that the syndrome does not depend on the codeword and, therefore, we do not require the NN to have a special structure, it can be arbitrary, but the best results were obtained with recurrent NNs \cite{Bennatan2018}. Later a syndrome-based approach was adapted to the transformer and denoising diffusion architectures \cite{ecct, diffusion}. We also note the papers (see, e.g. \cite{Jiang2020}) devoted to DNN-based code construction. For additional literature and a more detailed overview, we refer the reader to \cite{intro-review}. 


The papers above focus on the performance of hard-output decoding, i.e. the decoder is required to return the estimated information word. At the same time, modern receivers (such as MIMO receivers \cite{Studer-diss}) and modern codes consisting of short component codes \cite{Tanner81} require iterative (or turbo) decoders. Soft-output decoding is of critical importance for such schemes. We note that several papers (e.g. \cite{Bennatan2018, bp-cammerer-2022}) mention the possibility of obtaining a soft output by the proposed DNN architectures, but to the best of our knowledge, the quality of such output was not investigated in the literature. In what follows, we fill this gap.

Our contribution is as follows. We start with a syndrome-based approach \cite{Bennatan2018} and modify it to provide soft output. The major change is the training process and the loss function, including the regularization term, which controls the soft output quality. We demonstrate the performance of the new decoder for the $[64, 45]$ BCH code on the binary input AWGN channel. We choose such parameters as maximum a-posteriori (MAP) decoding is feasible for this code but has large complexity, which prevents the use of such a method in practice. Our decoder can be considered as an approximation of the MAP decoder with smaller computation complexity, in other words, we require our DNN to reproduce the MAP output. We discuss various regularization functions and compare the resulting distributions. Finally, to demonstrate soft output quality, we consider the iterative decoding scheme, namely the turbo product code (TPC) with $[64, 45]$ BCH codes as row and column codes. We show that the resulting DNN-based scheme is very close to MAP-based performance and significantly outperforms the Chase decoder-based solution~\cite{Chase1972} in combination with the soft output calculation \cite{cp-tpc}.


The paper is organized as follows. In section \ref{sec:soft-dec} the proposed preprocessing procedure described, NN model architecture, and soft-output quality metrics are introduced and applied for distribution optimization. The section ends with the decoding performance results and their discussion. Section \ref{sec:iterative-dec} provides a description of the proposed soft-decoding approach application in the TPC decoding scheme. The framework preprocessing steps for iterative decoding and model tuning steps are followed by a discussion of the results.

\section{Soft-input soft-output DNN-based decoding}
\label{sec:soft-dec}

\subsection{System model}

Let us describe the system model. The user aims to transmit a $k$-bit information word $\mathbf{u} \in \{0,1\}^k$. We assume the use of a binary linear block code $\mathcal{C}$ of length $n$ and dimension $k$. Let $\mathbf{H}$ and $\mathbf{G}$ denote parity-check and generator matrices of the code $\mathcal{C}$ accordingly. The information word $\mathbf{u}$ is first encoded into the codeword $\mathbf{c} = (c_1, \ldots, c_n) = \mathbf{u} \mathbf{G} \in \{0,1\}^n$. Then the binary phase-shift keying (BPSK) modulation is applied, implying the following mapping. 
\[
\mathbf{x} = \tau(\mathbf{c}), \quad \tau(\mathbf{c}) = (\tau(c_{1}), \ldots, \tau(c_{n})),
\]
where $\tau:\{0, 1\} \rightarrow \{1, -1\}$.

Modulated codeword $\mathbf{x}$ is transmitted over the AWGN channel, thus the receiver obtains corrupted codeword 
\[
\mathbf{y} = \mathbf{x} + \mathbf{z},
\]
where $\mathbf{y} = (y_1, \ldots, y_n) \in \mathbb{R}^{n}$, $\mathbf{z}\sim \mathcal{N}(0,\sigma^2 \mathbf{I}_n)$ and $\mathbf{I}_n$ is the identity matrix of size $n \times n$. In what follows by $E_s/N_0$ we denote signal-to-noise ratio, $E_s/N_0 = 1/2\sigma^2$.

As usual \cite{modern-codes}, the input of the decoder is presented as a vector $\bm{\gamma} = (\gamma_1, \ldots, \gamma_n)$ of log-likelihood ratios, where 
\begin{equation}
    \gamma_i = \log\frac{p(y_i | c_i = 0)}{p(y_i | c_i = 1)} = \frac{2 y_i}{\sigma^2}, \:\: i = 1, \ldots, n,
\end{equation}
where $\log(\cdot)$ stands for a natural logarithm and $p(x) = 1/\sqrt{2 \pi \sigma^2} \exp\left[-x^2/(2 \sigma^2) \right]$ is the probability density function of a random variable distributed as $\mathcal{N}(0,\sigma^2)$.

Now, let us describe the decoding performance metric. Let us start with hard output decoding, and let $\hat{\mathbf{u}} = (\hat{u}_1, \ldots, \hat{u}_k) \in \{0,1\}^k$ be the estimated information word. In what follows we utilize bit error rate (BER) $P_b = \frac{1}{k}\sum\nolimits_{i=1}^k \Pr[u_i \neq \hat{u}_i]$ and frame error rate (FER) $P_f = \Pr[\mathbf{u} \neq \hat{\mathbf{u}}]$.

To assess soft output quality, we compare the decoder output to the bit-wise MAP output $\bm{\gamma}^* = (\gamma_1^*, \ldots, \gamma_n^*)$, where for $i = 1, \ldots, n$ we have
\[
\gamma_i^* = \log \frac{ \Pr[c_i = 0 | \mathbf{y}]}{\Pr[c_i = 1 | \mathbf{y}]} = \log \frac{\sum\nolimits_{\mathbf{c} \in \mathcal{C}, c_i = 0} \exp\left[ (\mathbf{1} - \mathbf{c}) \bm{\gamma}^T \right]}{ \sum\nolimits_{\mathbf{c} \in \mathcal{C}, c_i = 1} \exp\left[ (\mathbf{1} - \mathbf{c}) \bm{\gamma}^T \right]},
\]
where $\mathbf{1}$ is the all-one vector, $\bm{\gamma}^T$ is the transpose of $\bm{\gamma}$. We refer the reader to \cite{richardson2008modern} for the derivation.

\subsection{Syndrome-based approach}

The proposed soft-output decoding framework inherits the syndrome-based structure described by Bennatan et al. \cite{Bennatan2018}. The original syndrome-based decoder implementation was designed for the system with multiplicative noise and its performance was discussed for hard decision decoding. In this paper, we propose to modify the pre- and postprocessing steps to adapt the framework for soft-output decoding in the AWGN channel.

In \cite{Bennatan2018} authors propose to pass vector $[|\mathbf{y}|,\mathbf{s}]$ as the input to the noise estimator, where $[\cdot,\cdot]$ denotes concatenation, $|\mathbf{y}|$ -- reliability vector and $\mathbf{s} = \text{bin}(\mathbf{y})\mathbf{H}^T$ -- binary syndrome. In such notation $\text{bin}(\cdot)$ implies a hard decision over the received vector. Instead, we propose to utilize the so-called \textit{soft syndrome} introduced by Lugosch et al. \cite{Lugosch2018} to avoid the hard-decision step in preprocessing. Since there is an isomorphism in between $(\{0,1\}, \oplus)$ and $(\{1,-1\}, *)$, the syndrome can be expressed as follows
\begin{equation}
    s_i = \prod_{j\in\mathcal{M}(i)}\text{sign}(y_j), \forall i \in [1,n-k]
\end{equation}
where $\mathcal{M}(i)$ is the set of columns in the $i$-th row of parity check matrix $\mathbf{H}$ equal to $1$.

Thus, the hard syndrome relaxation for input LLR vector $\bm{\gamma}$ can be introduced as
\begin{equation}
    \tilde{s}_i = \min_{j\in\mathcal{M}(i)} \lvert \gamma_j \rvert \prod_{j\in\mathcal{M}(i)}\text{sign}(\gamma_j), \forall i \in [1,n-k]
\end{equation}
For following description we denote noise estimator input vector by $\mathbf{d} = [|\bm{\gamma}|,\mathbf{\tilde{s}}]\in\mathbb{R}^{2n-k}$.

The proposed decoding algorithm is summarized in Algorithm~\ref{alg:soft-dec}. The noise estimation function is denoted by $\mathcal{F}$. In what follows $\mathcal{F}$ is chosen to be a DNN.

\begin{algorithm}[ht]
    \caption{Soft-output syndrome-based DNN decoding}
    \begin{algorithmic}[1]
    \label{alg:soft-dec}
    \renewcommand{\algorithmicrequire}{\textbf{Input:}}
    \renewcommand{\algorithmicensure}{\textbf{Output:}}
    \REQUIRE $\bm{\gamma}\in\mathbb{R}^{n}$ - input LLRs, $\mathbf{\tilde{s}}\in\mathbb{R}^{n-k}$ - soft syndrome
    \ENSURE  $\bm{\hat{\gamma}}\in\mathbb{R}^n$ - transmitted message LLRs estimation
    \STATE $\mathbf{\hat{z}} \gets \mathcal{F}([|\bm{\gamma}|,\mathbf{\tilde{s}}])$
    \STATE $\bm{\hat{\gamma}} \gets \bm{\gamma} - \text{sign}(\bm{\gamma})\odot\mathbf{\hat{z}}$
    \RETURN $\bm{\hat{\gamma}}$
    \end{algorithmic}
\end{algorithm}

\subsection{NN model architecture}

The main goal of the neural network $\mathcal{F}$ is to estimate the noise vector, and the choice of the best architecture remains an open question~\cite{Bennatan2018, ecct, diffusion}. In this paper, we focus on estimating the ability of the neural network framework to perform soft decoding. Based on the analysis of hard decoding quality and the time required to train different architectures, we choose a Stacked-GRU architecture~\cite{Bennatan2018}.

\begin{figure}[ht]
\centering
\includegraphics{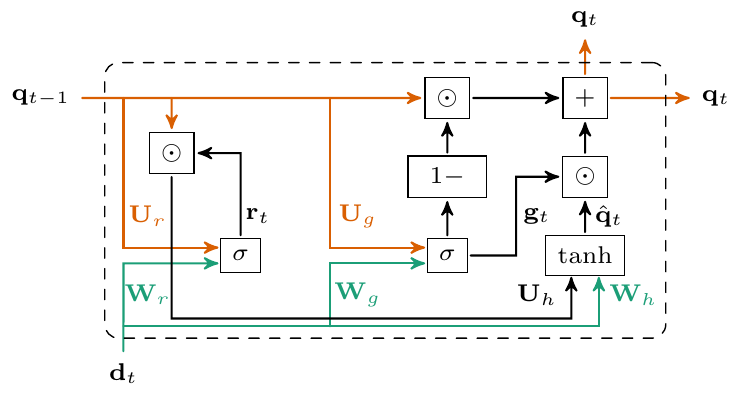}
\caption{Gated Recurrent Unit cell}
\label{fig:gru-cell}
\end{figure}

Stacked-GRU is a multi-layer Recurrent Neural Network (RNN) architecture composed of Gated Recurrent Unit (GRU) cells~\cite{gru} with trainable ``update'' and ``reset'' gates. Each GRU cell can be described by the following equations (see~\figurename~\ref{fig:gru-cell} for more details).
\begin{gather}
    \mathbf{g}_t = \sigma (\mathbf{W}_g\mathbf{d}_t + \mathbf{U}_g\mathbf{q}_{t-1} + \mathbf{b}_g),\\
    \mathbf{r}_t = \sigma (\mathbf{W}_r\mathbf{d}_t + \mathbf{U}_r\mathbf{q}_{t-1} + \mathbf{b}_r),\\
    \mathbf{\widehat{q}}_t = \tanh \big(\mathbf{W}_h\mathbf{d}_t + \mathbf{U}_h(\mathbf{r}_t\odot\mathbf{q}_{t-1})+\mathbf{b}_h\big),\\
    \mathbf{q}_t = \mathbf{g}_t\odot\mathbf{\widehat{q}}_t + (\mathbf{1}-\mathbf{g}_t)\odot\mathbf{q}_{t-1},
\end{gather}
where $\mathbf{d}_t$ is the input vector, $\mathbf{q}_t$ - output vector, $\mathbf{\widehat{q}}_t$ - candidate output vector, $\mathbf{g}_t$ - update gate vector, $\mathbf{r}_t$ - reset gate vector, $\mathbf{W},\mathbf{U}$ - trainable parameters matrices and $\mathbf{b}$ - trainable bias vectors. $\sigma(\cdot)$ denotes sigmoid function, $\tanh(\cdot)$ hyperbolic tangent and $\odot$ Hadamard product.
This architecture is widely used for Natural Language Processing (NLP).

To form a Stacked-GRU architecture, cells are bundled in two dimensions. The first stacking dimension is similar to the general fully connected NN (FCNN) layers. The output $\mathbf{q}_t$ of the preceding cell is passed to the feature input $\mathbf{d}_t$ of the subsequent cell. By $L$ we denote the total number of layers in the Stacked-GRU network. The second stacking dimension defines the recurrent structure of the network. The output of the preceding cells is passed as the hidden state $\mathbf{q}_{t-1}$ to the following cells. The initial hidden state of the network $\mathbf{q}_0$ is set to zero. We denote the total number of \textit{time steps} by $T$.

\begin{figure}[ht]
\centering
\includegraphics{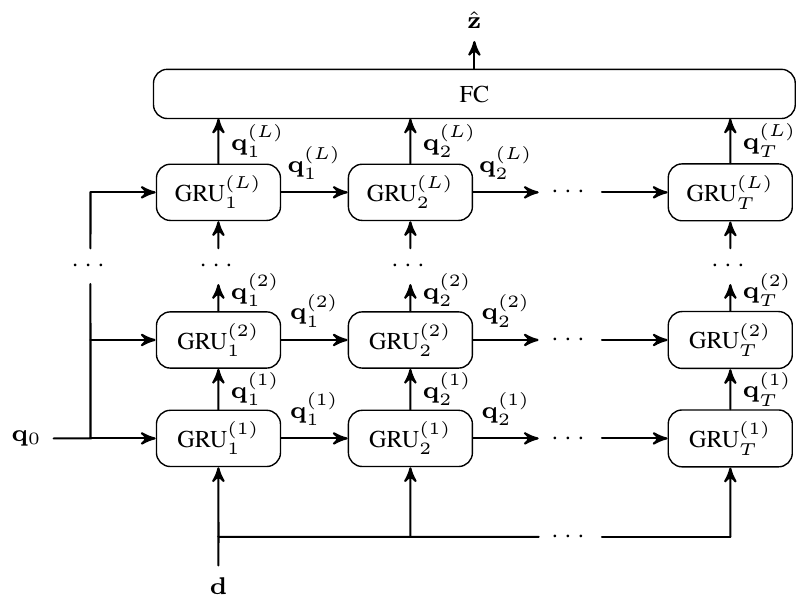}
\caption{Stacked-GRU model architecture}
\label{fig:stacked-gru}
\end{figure}

For a single input vector, Stacked-GRU NN generates $T$ vectors on the outputs of the last layer. We denote matrix of stacked output vectors by $\mathbf{Q}^{(L)} = [\mathbf{q}^{(L)}_1,\mathbf{q}^{(L)}_T]\in\mathbb{R}^{2n-k\times T}$, where superscript $L$ denotes index of the last layer and subscript $t$ denotes the GRU time step. In order to reduce the size of Stacked-GRU output, its vectorized representation $\text{vec}(\mathbf{Q}^{(L)})$ is passed to the single FC layer. The complete architecture of the noise estimator model is depicted in \figurename \ref{fig:stacked-gru}.

For model training, the loss is calculated from the framework soft output and binary codeword. The Binary Cross-Entropy (BCE) with sigmoid function is utilized.

\begin{equation}
\label{eq:bce-loss}
\begin{split}
    \mathcal{L}_{BCE}(\bm{\hat{\gamma}},\mathbf{c}) = -\frac{1}{n}\sum^n_{i=1}c_{i}\log\sigma(-\hat{\gamma}_{i})+ \\
    (1-c_{i})\log(1-\sigma(-\hat{\gamma}_{i}))    
\end{split}
\end{equation}

\subsection{Soft-output quality optimization}

\begin{table*}[t]
    \caption{Evaluation of the NN-decoder soft-output distribution\\ compared to MAP output with proposed regularizations}
    \centering
    \renewcommand{\arraystretch}{1.25}
    \begin{tabular}{|c|c|c|c|c|}
    \cline{2-5}
    \multicolumn{1}{c|}{}       & $\mathcal{L}_\text{BCE}$      & $\mathcal{L}_\text{BCE}+\alpha_\text{MSE}\mathcal{L}_\text{MSE}$  & $\mathcal{L}_\text{BCE}+\alpha_\text{KL}\mathcal{L}_\text{KL}$    & $\mathcal{L}_\text{BCE}+\alpha_\text{M}\mathcal{L}_\text{M}$  \\
    \cline{2-3}\hline
    Mean:   & 0.606     & 0.148         & 0.359         & 0.018     \\ 
    \hline
    Var:    & 11.752    & 7.170         & 8.539         & 0.231     \\ 
    \hline
    KL div.: &$3.158\cdot10^{-10}$ & $1.219\cdot10^{-10}$   & $1.181\cdot10^{-10}$  & $3.071\cdot10^{-10}$     \\ 
    \hline
    MSE:    & 4.906     & 3.547         & 4.759         & 4.864     \\ 
    \hline
    \end{tabular}
    \renewcommand{\arraystretch}{1}
    \label{table:reg-res}
\end{table*}

In order to optimize the soft-output distribution we propose to introduce a regularization term into the loss function for the last epochs of a model training procedure. Three types of regularization are proposed: Mean Squared Error (MSE), Kullback-Leibler (KL) divergence and moments-based. MAP decoder output LLRs $\bm{\gamma}^*$ are used as a reference.

MSE $\mathcal{L}_{MSE}$ and KL divergence $\mathcal{L}_{KL}$ regularizations are defined in a pointwise manner.
Moments-based regularization $\mathcal{L}_M$ is expressed as the weighted sum of the MSEs of the first and second moments of the decoder output distributions. 

\begin{equation}
    \mathcal{L}_{\text{MSE}}(\bm{\gamma}^*,\bm{\hat{\gamma}}) = \frac{1}{n}\sum_{i=1}^{n}(\gamma_i^* - \hat{\gamma}_i)^2
\end{equation}

\begin{equation}
    \mathcal{L}_{\text{KL}}(\bm{\gamma}^*,\bm{\hat{\gamma}}) = \sum_{i=1}^n \gamma_i^* \cdot \log\frac{\gamma_i^*}{\hat{\gamma}_i}
\end{equation}

\begin{equation}
\begin{split}
    \mathcal{L}_{\text{M}}(\bm{\gamma}^*,\bm{\hat{\gamma}}) = \rho_\text{M}\Big(\mathbb{E}(|\bm{\gamma}^*|) - \mathbb{E}(|\bm{\hat{\gamma}}|)\Big)^2 + \\
    (1-\rho_\text{M})\Big(\text{Var}(|\bm{\gamma}^*|) - \text{Var}(|\bm{\hat{\gamma}}|)\Big)^2
\end{split}
\end{equation}

The loss function with regularization term is expressed as
\begin{equation}
\mathcal{L} = \mathcal{L}_\text{BCE} + \alpha_\text{Reg} \mathcal{L}_\text{Reg}
\end{equation}
where $\mathcal{L}_{Reg}$ is the selected regularization metric and $\alpha_\text{Reg}$ its weight coefficient.

\begin{figure}[ht]
\centering
\includegraphics{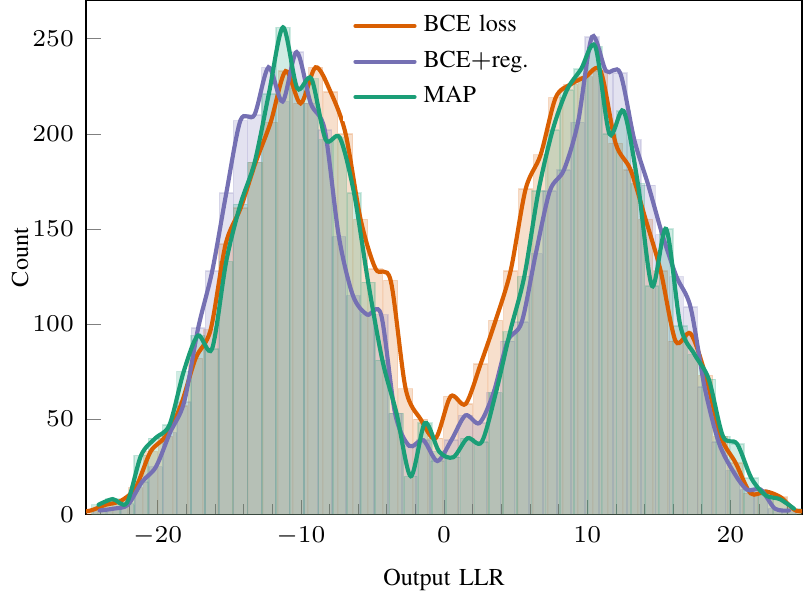}
\caption{Output LLR distributions histogram for $E_s/N_0 = 1 dB$ $[64,45]$ BCH code. The moments-based approach is used for regularization.}
\label{fig:reg-res}
\end{figure}

The results of the described regularization terms applied for the optimization of the soft output distribution of the NN decoder are summarized in Table \ref{table:reg-res} and depicted in \figurename\ref{fig:reg-res} for the moments-based approach. Table \ref{table:reg-res} evaluates the NN-decoder output distribution similarity to the MAP decoder in terms of metrics used for regularizations. Results are provided for NN-decoder trained on $[64, 45]$ BCH code with distribution evaluation on $E_s/N_0 = 1 dB$. Regularization terms weights were estimated empirically: $\alpha_{\text{MSE}}=0.01$, $\alpha_{\text{KL}}=10^{10}$, $\alpha_\text{M} = 0.1$, $\rho_\text{M}=0.95$.

\subsection{Simulation results}
\label{sec:soft-dec-exps}

To evaluate the decoding performance of the proposed framework, the Stacked-GRU model with a hidden size of $5n$, 4 layers, and 5 time-steps, as in \cite{Bennatan2018}, was trained on zero codewords with a batch size of $2^{13}$ codewords. The initial learning rate of Adam optimizer \cite{adam} was set to $10^{-3}$ with a further decrease to $10^{-6}$ by the ``reduce on plateau'' scheduler. Initial training was performed with BCE loss (\ref{eq:bce-loss}) only. MAP-based regularization terms were introduced for the last epochs only due to the high complexity of MAP decoding\footnote{we note that proposed soft-output DNN can be utilized without joint DNN-MAP fine tuning stage, if such is restricted by the complexity reasons.}.

The performance of soft-output DNN decoder was compared with the Chase decoder, Belief propagation with 50 decoding iterations and NN-Tanner \cite{5g-short-bp} with 20 decoding iterations.

\begin{figure}
\centering
\includegraphics{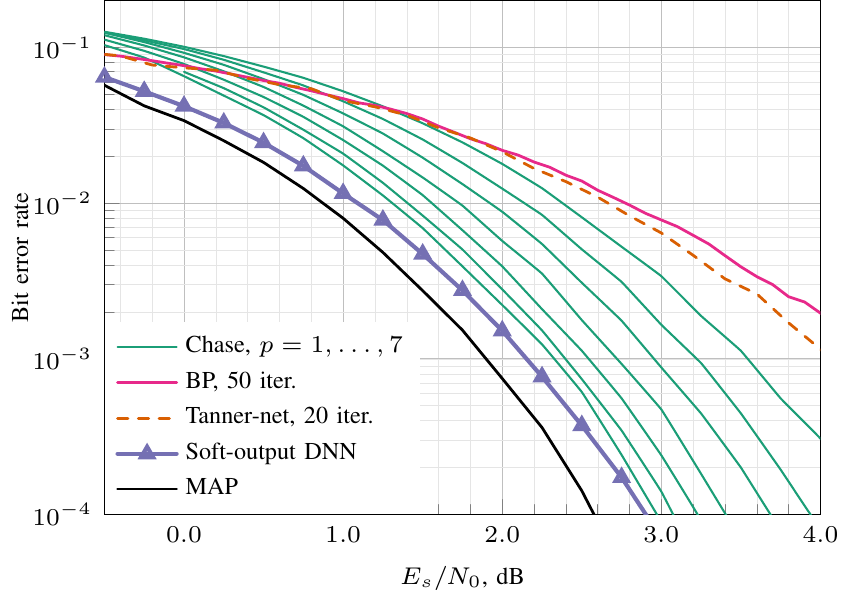}
\caption{Bit error rate results for $[64, 45]$ BCH code}
\label{fig:bch-dec}
\end{figure}

\section{NN iterative soft-output decoding}
\label{sec:iterative-dec}

To work in iterative decoding schemes, the decoder must be able to produce a soft output. Turbo Product Code (TPC) scheme was chosen to demonstrate the potential of using the proposed framework in iterative decoding schemes.

\subsection{Turbo product code}

\begin{figure}[tb]
\centering
\includegraphics{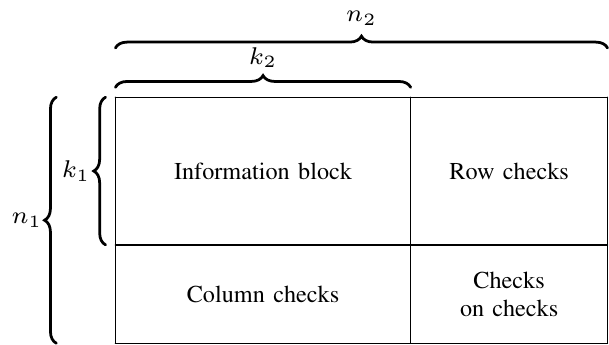}
\caption{TPC structure}
\label{fig:tpc-structure}
\end{figure}

Turbo Product Code (TPC) structure can be explained using the diagram in \figurename \ref{fig:tpc-structure}. TPC is constructed from two component codes in the systematic form with parameters $(n_1,k_1)$ and $(n_2,k_2)$ respectively. The encoding is performed in two steps. Initially, the information submatrix $k_1 \times k_2$ is encoded by the "column code" producing the "column checks" submatrix. Then the information and column check submatrices are encoded with the "row code", thus producing the "row checks" and "checks-on-checks" submatrices. The aggregate code rate of TPC is $R=(k_1 k_2)/(n_1 n_2)$ \cite{modern-codes}.

The iterative TPC decoding procedure is summarized in the algorithm \ref{alg:tpc-decoding}. There by $N$ we denote the number of decoding iterations, by $\mathcal{D}_c(\cdot), \mathcal{D}_r(\cdot)$ column and row decoding functions, by $\mathbf{L}_c\in\mathbb{R}^{n_2\times n_1}$, $\mathbf{L}_r\in\mathbb{R}^{n_1\times n_2}$ extrinsic information matrices and by $\alpha^{(i)}_c,\alpha^{(i)}_r\in[0,1]$ extrinsic LLRs scale factors on $i$-th iteration.

\begin{algorithm}[t]
    \caption{TPC decoding}
    \begin{algorithmic}[1]
    \label{alg:tpc-decoding}
    \renewcommand{\algorithmicrequire}{\textbf{Input:}}
    \renewcommand{\algorithmicensure}{\textbf{Output:}}
    \REQUIRE $\mathbf{\Gamma}\in\mathbb{R}^{n_1\times n_2}$ - channel output
    \ENSURE  $\mathbf{\widehat{\Gamma}}\in\mathbb{R}^{n_1\times n_2}$ - transmitted message estimation
    \STATE $\mathbf{L}_c, \mathbf{L}_r \gets 0$
    \STATE $\mathbf{\widehat{\Gamma}} \gets \mathbf{\Gamma}$
    \FOR {$i = 1$ to $N$}
    \item[] \textit{Column decoding}:
    \STATE $\mathbf{A} \gets \mathbf{\widehat{\Gamma}}^T - \mathbf{L}_c$
    \STATE $\mathbf{L}_0 \gets \mathcal{D}_c(\mathbf{A})$
    \STATE $\mathbf{L}_c \gets \alpha_c^{(i)}(\mathbf{L}_0 - \mathbf{A})$
    \STATE $\mathbf{\widehat{\Gamma}} \gets \mathbf{A}^T + \mathbf{L}_c^T$
    \item[] \textit{Row decoding}:
    \STATE $\mathbf{A} \gets \mathbf{\widehat{\Gamma}} - \mathbf{L}_r$
    \STATE $\mathbf{L}_0 \gets \mathcal{D}_r(\mathbf{A})$
    \STATE $\mathbf{L}_r \gets \alpha_r^{(i)}(\mathbf{L}_0 - \mathbf{A})$
    \STATE $\mathbf{\widehat{\Gamma}} \gets \mathbf{A} + \mathbf{L}_r$
    \ENDFOR
    \RETURN $\mathbf{\widehat{\Gamma}}$
    \end{algorithmic}
\end{algorithm}

\subsection{NN TPC decoding}
To utilize the soft-output DNN decoder we pre-train the model for component code decoding, as described in Section \ref{sec:soft-dec}, and then fine-tune it in the iterative scheme. 

One of the advantages of a syndrome-based approach lies in its robustness to overfitting. The noise estimation NN-model is trained on reliability vectors and syndromes, which do not depend on the transmitted codeword. Thus, the model can be trained on a zero codeword with different realizations of noise. However, the performance of the model trained on the defined range of Signal-to-Noise Ratios degrades for values over the range. This problem arises in the iterative schemes, since with each iteration the absolute value of the output LLRs grows. The proposed iterative decoding approach does not require training a separate model for each decoding iteration. The same pretrained DNN decoder is utilized for all TPC iterations. To solve the issue of growing LLRs we apply $L^1$ batch normalization for the input of the decoding framework.

In the NN-TPC decoding scheme extrinsic LLRs scale factors $\alpha_c, \alpha_r$ are initialized as the trainable parameters, thus during the fine-tuning stage their optimal value is calculated by the gradient descent jointly with the decoding model.

The loss function for the NN-TPC fine-tuning stage is the exponentially weighted sum of BCE loss (\ref{eq:bce-loss}) of all decoding iterations.

\begin{gather}
    \bm{\beta} = [e^0,\dots,e^{2N-1}] \\
    \mathcal{L}_\text{NN-TPC} = \frac{1}{2N\|\bm{\beta}\|_1} \sum_{j=1}^{2N} \beta_j\mathcal{L}_{BCE}(\mathbf{\widehat{\Gamma}}_j,\mathbf{C})
\end{gather}

Where $\mathbf{\widehat{\Gamma}}_j$ is decoded by columns/rows message LLRs on iteration $\lceil j/2\rceil$ and $\mathbf{C}$ is transmitted binary TPC message.

\subsection{Simulation results}

To evaluate the proposed framework in an iterative decoding scheme, we use TPC with $[64,45]$ BCH as component code. Soft-output DNN model was initially trained on component code, as described in \ref{sec:soft-dec-exps}. Then the model was fine-tuned in the TPC decoding scheme with a learning rate $10^{-6}$ for 4000 epochs with a batch size of $256$. The extrinsic scales $\alpha_c,\alpha_r$ were initialized by the value $0.7$.

The performance of the soft-output DNN decoder is compared to the Chase-Pyndiah algorithm \cite{cp-tpc} for $N=2$ and $N=4$ TPC decoding iterations. Chase-Pyndiah results were obtained with the AFF3CT toolbox \cite{aff3ct}.

\begin{figure}
\centering
\includegraphics{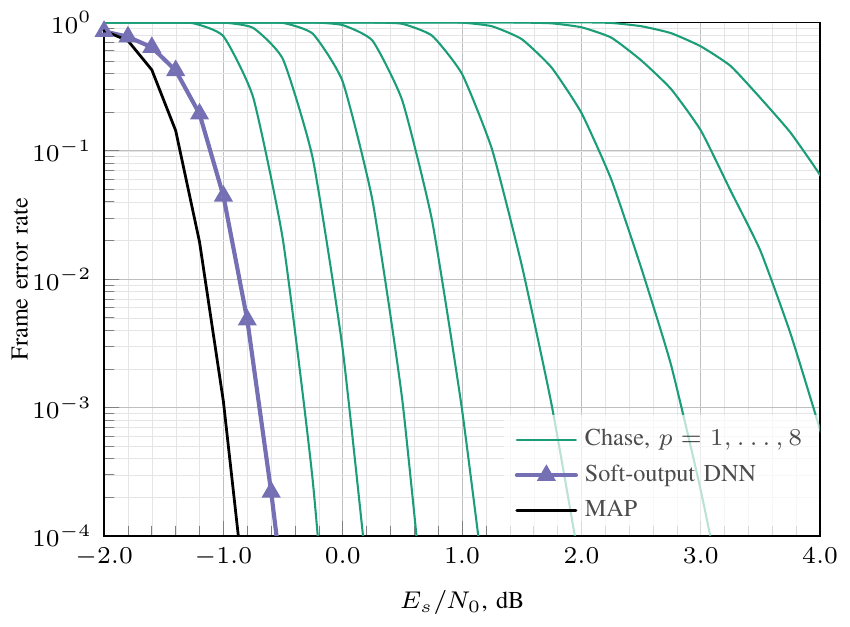}
\caption{FER results for TPC decoding scheme with BCH(64,45) as component code, 2 iterations}
\label{fig:tpc-2-iter}
\end{figure}

\begin{figure}
\centering
\includegraphics{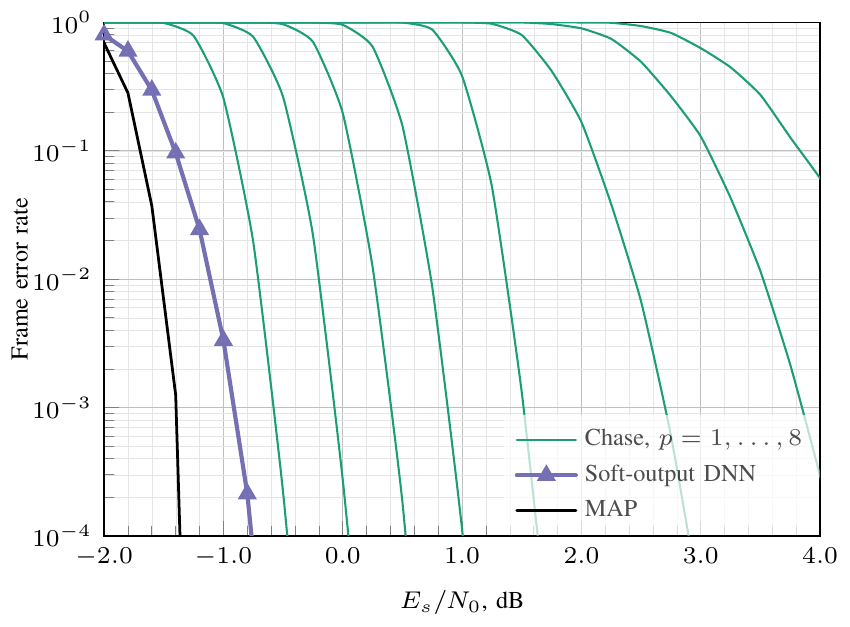}
\caption{FER results for TPC decoding scheme with BCH(64,45) as component code, 4 iterations.}
\label{fig:tpc-4-iter}
\end{figure}

Further research directions are as follows. In the real-world applications with increasing requirements to latency, memory and power usage, there is a challenge of decoder complexity reduction without loss of decoding performance. DNN utilized for noise estimation in this paper has relatively high complexity ($2.2\cdot 10^6$ parameters). The question of optimal architecture selection for soft-output decoding is an open question. Apart from that, we point out the model weights adaptive quantization \cite{quant-thesis,quant-1,quant-2}, activation functions approximation \cite{quant-2} and model weights pruning \cite{pruning} as the potential directions of soft-output DNN complexity reduction.

\bibliographystyle{IEEEtran}
\bibliography{refs}

\end{document}